\documentclass[a4paper,fleqn,usenatbib]{mnras}
\usepackage{newtxtext,newtxmath,hyperref}
\usepackage[T1]{fontenc}
\usepackage{ae,aecompl}
\usepackage[dvipdfmx]{graphicx}	
\usepackage{amsmath}	
\usepackage{amssymb}	
\usepackage{multirow,cases,colortbl}
\usepackage{url}
\usepackage{bm}
\usepackage{lscape}
\usepackage{ulem,xcolor}

\newcommand{\Msun}{\ensuremath{\mathrm{M}_\odot}}

\title[The origin of hotspots around Sgr A*]{The origin of hotspots around Sgr A*: Orbital or pattern motion?}

\author[T. Matsumoto et al.]{Tatsuya Matsumoto$^{1,2,3}$\thanks{E-mail: tatsuya.matsumoto@mail.huji.ac.il}\thanks{JSPS Research Fellow}, Chi-Ho Chan$^{1,4}$, and Tsvi Piran$^{1}$
\\
$^{1}$Racah Institute of Physics, Hebrew University, Jerusalem, 91904, Israel\\
$^{2}$Research Center for the Early Universe, Graduate School of Science, University of Tokyo, Tokyo 113-0033, Japan\\
$^{3}$Department of Physics, Graduate School of Science, University of Tokyo, Tokyo 113-0033, Japan\\
$^{4}$School of Physics and Astronomy, Tel Aviv University, Tel Aviv 69978, Israel
}

\pubyear{2020}

\begin{document}
\label{firstpage}
\pagerange{\pageref{firstpage}--\pageref{lastpage}}
\maketitle

\begin{abstract}
The Gravity Collaboration detected a near-infrared hotspot moving around Sgr A* during the 2018 July 22 flare.
They fitted the partial loop the hotspot made on the sky with a circular Keplerian orbit of radius $\simeq7.5\,r_{\rm g}$ around the supermassive black hole (BH), where $r_{\rm g}$ is the gravitational radius.
However, because the hotspot traversed the loop in a short time, models in which the hotspot tracks the motion of some fluid element tend to produce a best-fit trajectory smaller than the observed loop. 
This is true for a circular Keplerian orbit, even when BH spin is accounted for, and for motion along a RIAF streamline.
A marginally bound geodesic suffers from the same problem; in addition, it is not clear what the origin of an object following the geodesic would be.
The observed hotspot motion is more likely a pattern motion.
Circular motion with $r\simeq12.5\,r_\mathrm g$ and a super-Keplerian speed $\simeq0.8\,c$ is a good fit.
Such motion must be pattern motion because it cannot be explained by physical forces. The pattern speed is compatible with magnetohydrodynamic perturbations, provided that the magnetic field is sufficiently strong.
Circular pattern motion of radius $\sim20\, r_{\rm g}$ on a plane above the BH is an equally good alternative; in this case, the hotspot may be caused by a precessing outflow interacting with a surrounding disk.
As all our fits have relatively large radii, we cannot constrain the BH spin using these observations.
\end{abstract}


\section{Introduction}
Decades-long monitoring of stellar orbits within sub-arcseconds of Sgr A*, the radio source at the Galactic Center, revealed the presence of a supermassive black hole 
\citetext{BH; see \citealp{Genzel+2010} for a review} with mass $M\simeq4.15\times10^{6}\,\Msun$ \citep{Schodel+2002,Ghez+2003,Gillessen+2009,Abuter+2019}.
The source Sgr A* is unusually dim: its bolometric luminosity $\sim10^{36}\,\rm erg\,s^{-1}$ is much smaller than its Eddington luminosity $\sim10^{44}\,\rm erg\,s^{-1}$.
The faintness means that the accretion flow around Sgr A* cannot be a radiatively efficient disk \citep{Shakura&Sunyaev1973}, but could instead be a radiatively inefficient accretion flow \citetext{RIAF; see \citealp{Yuan&Narayan2014} for a review}.
RIAF models reproduce the emission of Sgr A* in the quiescent (or steady) state \citep{Narayan+1995,Narayan+1998,Yuan+2003,Moscibrodzka+2009,Moscibrodzka+2014,Dexter+2020}.

One puzzle concerning Sgr A* is the origin of its near-infrared (NIR, \citealt{Genzel+2003}) and X-ray \citep{Baganoff+2001} flares. 
NIR flares occur $\sim4$ times per day and last for $\sim\,\rm hr$.
Their luminosity rises from $\sim10^{34}\,\rm erg\,s^{-1}$ at quiescence to $\sim10^{35}\,\rm erg\,s^{-1}$ at peak.
NIR flares are also characterized by their large ($\sim40\,\%$) linear polarization and temporally changing polarization angle \citep{Eckart+2006,Meyer+2006,Trippe+2007}.
In addition, a quasi-periodic substructure of $\simeq20\,\rm min$ \citep{Genzel+2003,Trippe+2007,DoddsEden+2009} and very short time-variability of $\lesssim50\,\rm s$ \citep{DoddsEden+2009} are detected.
X-ray flares show a larger increase in luminosity, from $\sim10^{33}\,\rm erg\,s^{-1}$ at quiescence to $\sim10^{35}\,\rm erg\,s^{-1}$ at peak.
Every X-ray flare is accompanied by an NIR flare but not vice versa \citep{Hornstein+2007}.

Several flare models have been proposed.
Although the X-ray emission mechanism is still debatable \citep{Markoff+2001,Yuan+2004,YusefZadeh+2006,DoddsEden+2009,DoddsEden+2010,Sabha+2010,Kusunose&Takahara2011,Ponti+2017}, the strongly linearly polarized NIR light could be synchrotron emission from non-thermal particles. These particles may be accelerated in turbulence, shocks, or magnetic reconnection \citep{DoddsEden+2010}.
The quasi-periodic structure and rapid variability of NIR flares suggest that the emission region is compact and close to the BH \citep{Broderick&Loeb2005,Broderick&Loeb2006b,Hamaus+2009,Vincent+2014}.

Recently, \citet[hereafter G18]{Abuter+2018b} carried out $K$-band interferometric observations of NIR flares using the GRAVITY instrument on the Very Large Telescope. They measured the motion of the flux centroid of the hotspot  with an astrometric accuracy of $\sim10\,\muup\mathrm{as}$ \citep[corresponding to $2 \,r_\mathrm g$,][]{Abuter+2017}.
G18 and \citet[hereafter G20]{Baubock+2020} fitted the hotspot trajectory with circular Keplerian orbits, finding best-fit radii of $7-10\,r_{\rm g}$ for the different flares, where $r_{\rm g}=GM/c^2$ is the gravitational radius, $G$ is the gravitational constant, and $c$ is the speed of light.
However, their best-fit orbits are entirely inside of the observed hotspot locations (see Fig.~\ref{fig radius}).
This motivates us to consider other models for the hotspot trajectory.

In this paper, we fit the hotspot trajectory with several kinematic models: circular Keplerian, geodesic, radiatively inefficient accretion flow (RIAF), super-Keplerian pattern (a hotspot moving faster than Keplerian along a circular trajectory), and precessing pattern.
In \S \ref{sec obs summary}, we summarize the observations and analysis by G18.
We describe our fitting method in \S \ref{sec method}.
In \S \ref{sec model}, we discuss the best-fit results for our various models.
We present our discussion and summary in \S \ref{sec summary}.
We adopt a distance of $d=8.18\,\rm kpc$ to Sgr A* \citep{Abuter+2019}; for a BH of mass $M=4.15\times10^6\,\Msun$, the gravitational radius spans an angle of $r_{\rm g}/d=5.01\,\muup\mathrm{as}$ on the sky at this distance.
Light crosses the gravitational radius in $r_{\rm g}/c=20.5\,\rm s$.

\section{Summary of GRAVITY observations}\label{sec obs summary}
G18 observed three NIR flares of Sgr A* on 2018 May 27, July 22, and July 28.
During the July 22 flare, the hotspot made a partial loop around a center consistent with the location of the BH as determined by monitoring the orbital motion of S2 (G18).
The duration of the flare and the angular extent of the partial loop are $\simeq30{\,\rm min}\simeq90\,(r_{\rm g}/c)$ and $\simeq100-150{\,\muup\mathrm{as}}\simeq20-30\,(r_{\rm g}/d)$, respectively.
The other two flares have similar duration, but the hotspot trajectories do not make a discernible loop.
The linear polarization angle of the July 28 flare rotated through $180^\circ$ over the course of the flare, which G18 explained with a poloidal magnetic field around the BH.

Hereafter we focus on the July 22 flare that was extensively analysed by the Gravity Collaboration; this allows our results to be compared directly to theirs. G18 fitted the hotspot trajectory with a circular Keplerian orbit, assuming that the orbit is centered on the two-dimensional median of the observed hotspot locations and taking into account light bending.
The best-fit radius and inclination are $r=7\,r_{\rm g}$ and $i=160^\circ$, respectively (see the first row of Table~\ref{table result} for the other fit parameters, and Fig.~\ref{fig fit} for a depiction of the fit).
G20 performed a similar fit but allowed the center of the orbit to vary within the error box of the BH location constrained by S2.
The best-fit parameters are $r=8.5\,r_{\rm g}$ and $i=155^\circ$. 

\begin{figure}
\begin{center}
\includegraphics[width=85mm, angle=0]{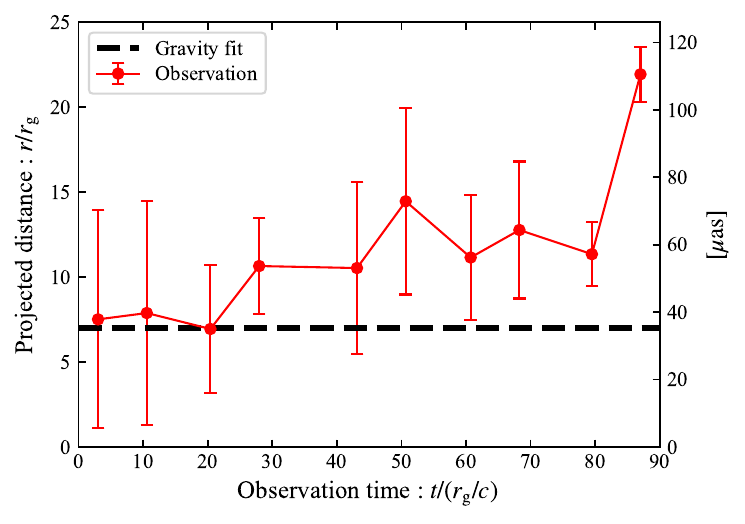}
\caption{Projected distances from the hotspot locations during the 2018 July 22 flare to the BH location determined by the G18 fit.
All hotspot locations are outside the radius $r=7\,r_{\rm g}$ of the best-fit circular Keplerian orbit.}
\label{fig radius}
\end{center}
\end{figure}

As already remarked by G20, both best-fit orbits lie entirely interior of the observed hotspot locations.
In Fig.~\ref{fig radius}, we plot the projected distances from the observed hotspot locations to the BH location determined by the G18 fit.
For comparison, we include the radius of the best-fit G18 orbit; we show it as a horizontal line because the best-fit orbit, with an inclination of $160^\circ$, is a circle on the sky to within $3\%$.
If the best-fit orbit were to pass through the observed hotspot locations, the points should be distributed randomly around the horizontal line; this is clearly not the case.
If the astrometric error is Gaussian, the probability that all 10 observed hotspot locations lie outside the best-fit orbit is merely $(1/2)^{10}\sim10^{-3}$. 
The best-fit G20 orbit has a larger radius of $8.5\,r_{\rm g}$, but it still lies inside the observed hotspot locations (see their Fig.~2). We cannot produce a figure analogous to Fig.~\ref{fig radius} for this fit because the best-fit BH location is not provided.

The problem with fitting circular Keplerian orbits (considered by G18 and G20) to the observed hotspot locations is that the radius $r$ of a circular Keplerian orbit in Schwarzschild spacetime is tied to its period $P_\mathrm K\propto r^{3/2}$. A particle completing three quarters of an orbit within $\simeq30\,\mathrm{min}$, as the observations seem to suggest, must be at $r\simeq7\,r_{\mathrm g}\,(P_{\rm K}/40\,\rm min)^{2/3}$, which is smaller than the spread of the observed hotspot locations on the sky. Circular Keplerian fits may sacrifice the best-fit radius in an attempt to reproduce the observed orbital time. The unsatisfactory nature of these fits motivates us to seek out models that break the degeneracy between radius and orbital time.

\section{Methods}\label{sec method}

\begin{figure}
\begin{center}
\includegraphics[width=85mm, angle=0]{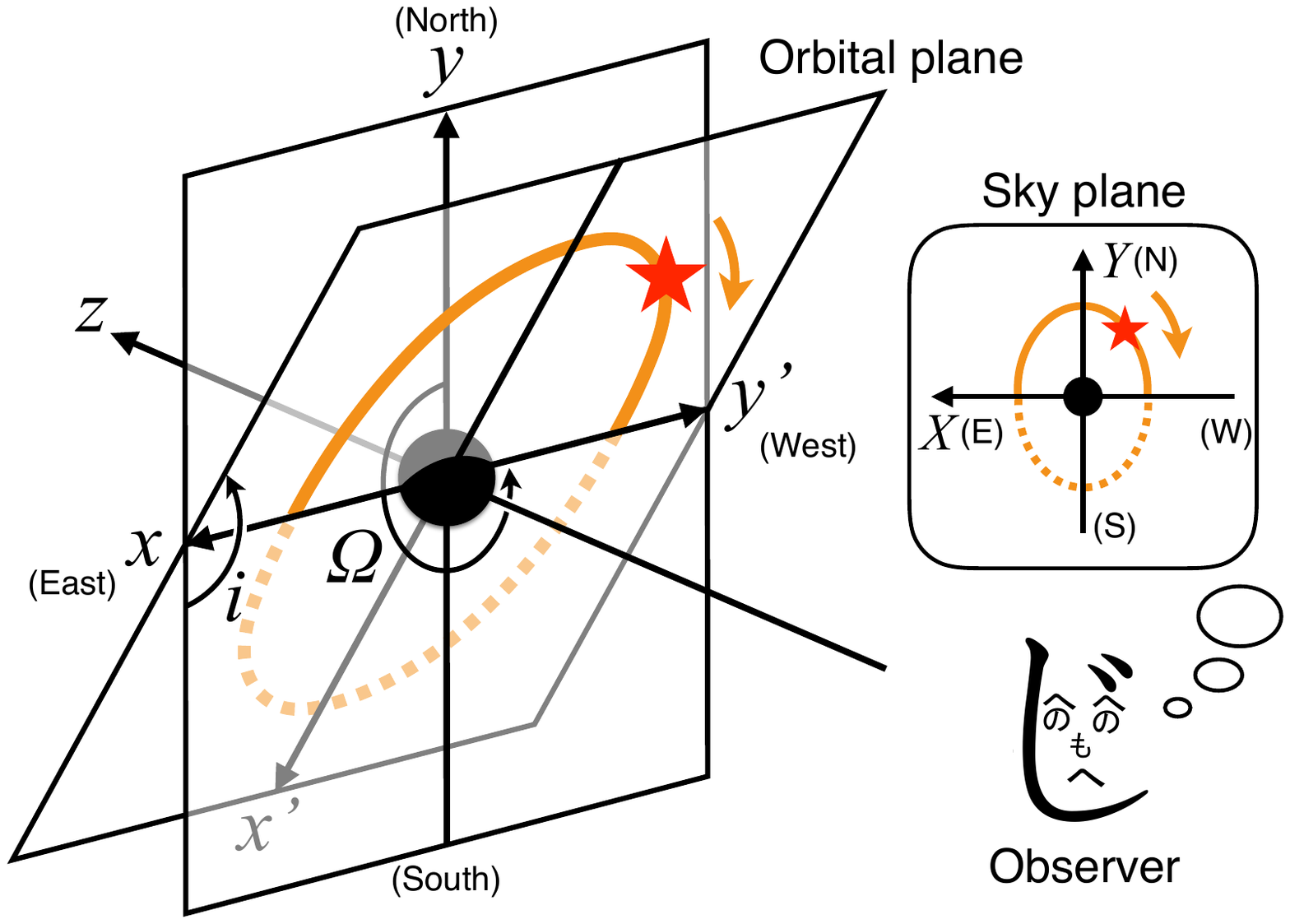}
\caption{Coordinate systems used in this paper. The longitude of the ascending node $\Omega$ and the inclination $i$ are marked with black counterclockwise arrows.
We chose $\Omega=270^\circ$ and $i>90^\circ$ in this figure, which results in the hotspot (red star) rotating clockwise around the BH from the point of view of the observer.
The hotspot trajectory on the sky is modified by light bending, but that effect is not included in this schematic diagram.
The part of the trajectory due to light emitted when the hotspot is at $z<0$ is shown with a solid curve, and at $z>0$ with a dashed curve; light from the latter part suffers greater deflection. To better depict the three-dimensional setting, we mark in gray objects that are behind the orbital plane.}
\label{fig coordinate}
\end{center}
\end{figure}

Fig.~\ref{fig coordinate} defines the coordinate systems we use \citetext{see \citealp{Paumard+2006} for details}.
The three-dimensional hotspot orbit is described in the Cartesian coordinate system $(x,y,z)$. The origin of the coordinate system coincides with the BH, the positive $x$- and $y$-axes are parallel to the sky plane, increasing toward the east and the north respectively, and the positive $z$-axis points directly away from the observer.

We consider only planar orbits for simplicity. We erect another Cartesian coordinate system $(x',y',z')$ such that the $x'$-$y'$ plane is the orbital plane and the positive $y'$-axis contains the ascending node, that is, the point where the hotspot moves across the $x$-$y$ plane away from the observer. The longitude of the ascending node $\Omega$ is the angle around the positive $z$-axis from the positive $y$-axis (north) to the positive $y'$-axis. The inclination angle $i$ is the angle between the positive $z$- and $z'$-axes. The relative orientation of the $x$-$y$ and $x'$-$y'$ planes is fully determined by $(\Omega,i)$.
The coordinate systems of the two planes are related by
\begin{align}
x&=x^\prime \cos i \cos\Omega+y^\prime \sin\Omega,
   \label{eq x}\\
y&=-x^\prime \cos i \sin\Omega+y^\prime \cos\Omega.
   \label{eq y}
\end{align}
Orbits are defined to rotate clockwise on the $x'$-$y'$ plane as viewed from the positive $z'$-axis; therefore, the observer sees the hotspot rotating clockwise (counterclockwise) on the sky if $90^\circ<i\leq180^\circ$ ($0\leq i<90^\circ$).

We describe the hotspot location on the sky with another coordinate system $(X,Y)$. The coordinates $X$ and $Y$ increase toward the east and the north, respectively.
The origin of the coordinate system is the two-dimensional median of the observed hotspot locations; note that the error box of the BH location constrained by S2, $(X_\mathrm{BH},\,Y_\mathrm{BH})=(30\pm50\,\muup\mathrm{as},\,-50\pm50\,\muup\mathrm{as})$, includes the origin. We assume the hotspot is an isotropic point source, moving as prescribed by our models in \S\ref{sec model}. To include light bending, we trace light rays from each emitting location $(x,y,z)$ of the hotspot to the observed location $(X,Y)$ on the sky assuming the BH has no spin \citep{Cunningham&Bardeen1973,Luminet1979,Broderick&Loeb2005,Broderick&Loeb2006b,Hamaus+2009}. This is performed with the open-source code \texttt{geokerr} \citep{Dexter&Agol2009}.

The goodness of fit is quantified by
\begin{align}
\chi^2=\sum_{i=1}^{N}\biggl[\Big(\frac{X(t_i)-X_i}{\sigma_{X_i}}\Big)^2+\Big(\frac{Y(t_i)-Y_i}{\sigma_{Y_i}}\Big)^2\biggl],
\end{align}
where $N$ is the number of observations ($N=10$ for the July 22 flare), $t_i$ is the $i$th observation time, and $\big(X_i\pm\sigma_{X_{i}},\,Y_i\pm\sigma_{Y_i}\big)$ is the $i$th observed hotspot location and its error bars.
The reduced $\chi^2$ is $\chi^2_{\rm r}\equiv\chi^2/(2N-N_\mathrm f)$, where $N_\mathrm f$ is the number of free parameters of the model.
The value of $N_\mathrm f$ varies from model to model and is between six and eight in our models. 
All models share six parameters: two $(\Omega,i)$ for the orientation of the orbital plane, two for the initial position of the hotspot on the $x'$-$y'$ plane, and two $(X_\mathrm{BH},Y_\mathrm{BH})$ for the BH location on the sky.

In \S\ref{sec precession model}, we consider a model in which the hotspot is produced, for example, when a precessing outflow interacts with a surrounding disk. We assume for simplicity that the hotspot traces out in the $(x,y,z)$ coordinate system a circle whose center is offset from the BH.
Because we assume that the hotspot trajectory is planar in this model, we describe the hotspot trajectory using the formalism laid out above. The only modification is that the origin of the $(x,y,z)$ coordinate system is not at the BH, but at the center of the hotspot circle.

\section{Models and fitting results}
\label{sec model}

\begin{figure*}
\begin{center}
\includegraphics[width=170mm, angle=0]{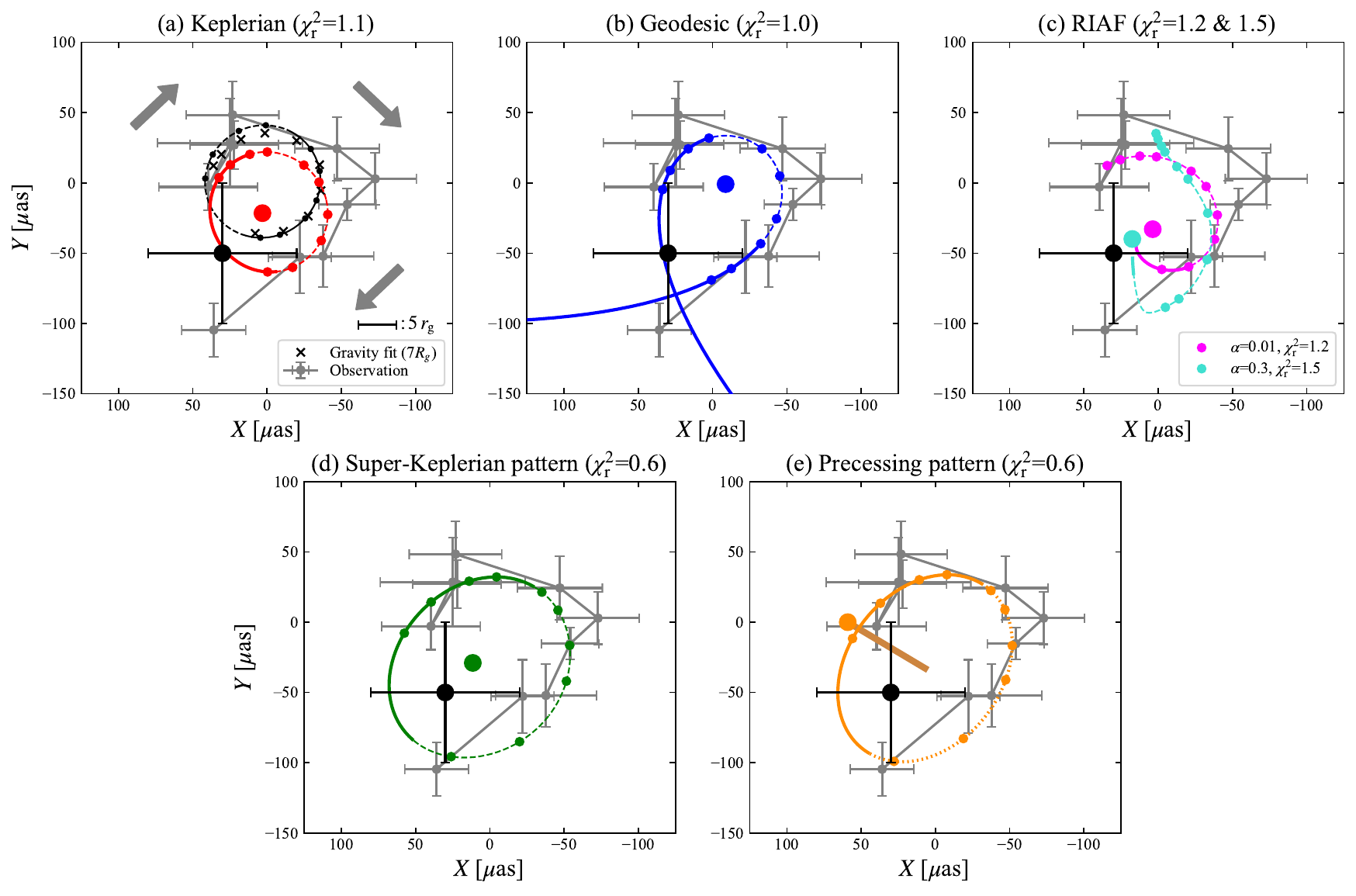}
\caption{Hotspot trajectories on the sky for different kinematic models of the 2018 July 22 flare; see Table \ref{table result} for the best-fit parameters.
The gray crosses show the locations of the hotspot at the ten times when it was observed, and the origin of the coordinate system is their two-dimensional median.
The black cross is the $1\sigma$ error bars on the location of the BH constrained by the orbital motion of S2 (G18).
The colored points mark the locations on the best-fit trajectories at each time the hotspot was observed.
In Panels (a)--(d), the best-fit BH location is shown by a large colored dot. The solid (dashed) part of the best-fit trajectories is due to light emitted when the hotspot is in front of (behind) the BH (see Fig.~\ref{fig coordinate}).
In Panel (a), the small black crosses are the hotspot locations of the best-fit G18 model. The small black dots connected by a black curve show our reproduction of that fit using the information provided by G18. In Panel (e), the elliptical trajectory is produced by a hotspot moving along a circle on a plane inclined to our line of sight and not passing through the BH. The half of the circle closer to (further away from) the observer than its center is drawn as a solid (dotted) curve. The BH can be anywhere along an axis going perpendicularly through the center of the circle. This axis is shown as a brown line, cropped to the error box of the location of Sgr A*, and further cropped subject to the condition that the BH is behind the hotspot circle.
The large yellow dot denotes the possible BH location furthest away from the center of the circle.}
\label{fig fit}
\end{center}
\end{figure*}

We consider several kinematic models of hotspot motion.
We divide the models into two types: material models, in which the hotspot is produced by an emitting fluid element;
and pattern models, in which the motion of the hotspot is a pattern motion.
The best-fit models and their parameters are shown in Fig.~\ref{fig fit} and Table \ref{table result}, respectively.

\begin{table*}
\begin{center}
\caption{Best-fit parameters for different kinematic models of the 2018 July 22 flare; see Fig.~\ref{fig fit} for the hotspot trajectories on the sky.}
\label{table result}
\begin{tabular}{lrrrrrlr}
\hline
Model &Initial&Initial&Longitude of&Inclination&BH location&Additional&Reduced $\chi^2$\\
&radius&argument&ascending node&&on sky plane&model parameters\\
&&of latitude&&&&\\
&$r_0/r_{\rm g}$&$\varphi_0$ [deg]&$\Omega$ [deg]&$i$ [deg]&$(X_{\rm BH}/r_\mathrm g,Y_{\rm BH}/r_\mathrm g)$&&$\chi^2/{\rm dof}=\chi^2_{\rm r}$\\
\hline\hline
G18 (circular Keplerian)&7&--&160&160&$(0,0)$&--&1.2\\
G20 (circular Keplerian)&8.5&--&--&155&--&--&1.6\\
\hline
Circular Keplerian&7 (fixed)&345.2&160 (fixed)&160 (fixed)&$(0,0)$ (fixed)&--&$23.6/14=1.7$\\
Circular Keplerian&7.5&229.8&12.5&152&$(3.1,-21.5)$&--&$15.4/14=1.1$\\
\multirow{2}{*}{Geodesic}&\multirow{2}{*}{8.3}&\multirow{2}{*}{198.9}&\multirow{2}{*}{25.1}&\multirow{2}{*}{162}&\multirow{2}{*}{$(-8.7,-0.9)$}&$E^a=1({\rm fixed})$&\multirow{2}{*}{$11.8/12=1.0$}\\
&&&&&&${l/(r_{\rm g}c)}=4.2$\\
RIAF ($\alpha=0.01$)&9.8&276.0&39.6&155&$(3.7,-32.9)$&--&$17.4/14=1.2$\\
RIAF ($\alpha=0.3$)&22.8& 337.2& 361.9&96&$(17.4,-39.9)$&--&$21.3/14=1.5$\\
\hline
\multirow{2}{*}{Super-Keplerian pattern}&\multirow{2}{*}{12.5}&\multirow{2}{*}{170.7}&\multirow{2}{*}{323.5}&\multirow{2}{*}{141}&\multirow{2}{*}{$(11.5,-29.0)$}&${v_{\varphi}}^b=0.76\, c$&\multirow{2}{*}{$7.8/13=0.6$}\\
&&&&&&$\,\,\,\,\,$(or ${\omega/\omega_{\rm K}}^c=2.7$)\\

Precessing pattern&14.1&173.4&327.8&138&$(7.0,-32.8)^d$&${v_{\varphi}}^b=0.88\, c$&$7.6/13=0.6$\\
\hline 
\multicolumn{8}{l}{$^a$ specific energy, $^b$  rotational velocity, $^c$ ratio of angular velocity to Keplerian orbital frequency,}\\
\multicolumn{8}{l}{$^d$ not the BH location on the sky plane, but the center of the circular pattern motion projected on the sky}
\end{tabular}
\end{center}
\end{table*}

\subsection{Material models}
\label{sec material model}

\subsubsection{Circular Keplerian orbit}
\label{sec kepler}

We fit a circular Keplerian orbit to the observations and check that we reproduce the results of G18 and G20.
The hotspot orbit in this model is
\begin{align}
x^\prime(t)&=r_0\cos(\omega_{\rm K} t+\varphi_0),
   \label{eq kepler x}\\
y^\prime(t)&=-r_0\sin(\omega_{\rm K} t+\varphi_0),
   \label{eq kepler y}
\end{align}
where $r_0$ and $\varphi_0$ are the radius and argument of latitude of the hotspot at $t=0$, respectively, and $\omega_{\rm K}=(c/r_{\rm g})(r_0/r_{\rm g})^{-3/2}$ is the Keplerian orbital frequency at $r=r_0$ in Schwarzschild spacetime.
The minus sign in Eq. \eqref{eq kepler y} reflects our convention of clockwise hotspot motion on the orbital plane (see Fig.~\ref{fig coordinate}).
This model is fully determined by the six parameters common to all models.

First, we fix all but one parameter to the G18 values and fit for the only parameter not given, $\varphi_0$.
Panel (a) of Fig.~\ref{fig fit} shows that we can reasonably reproduce the G18 result.
The agreement is mediocre over the part of the orbit on the far side of the BH ($z>0$, dashed black curve).
This may be because G18 assumed the hotspot has a finite size, while we assume the hotspot is a point source; strong lensing may cause the flux centroids of these two kinds of hotspots to be misaligned.
Second, we perform a fit in which all six parameters vary freely.
The red curve in Panel (a) is the result.
We find $r_0=7.5\,r_{\rm g}$ and $i=152^\circ$, comparable to the G20 results of $r_0=8.5\,r_{\rm g}$ and $i=155^\circ$. The G20 fit is not plotted because not all best-fit parameters are given.

G18, G20, and we all found best-fit orbits that lie entirely inside of the observed hotspot locations.
This is because of a fundamental tension between the observed hotspot making a relatively large loop on the sky (three quarters of a circle with radius $\gtrsim7-8.5\,r_{\rm g}$) and it doing so within a relatively short time (30 min).
Circular Keplerian fits may be more strongly driven by the latter constraint and thus pushed toward smaller $r_0$. 

\begin{figure}
\begin{center}
\includegraphics[width=60mm, angle=0]{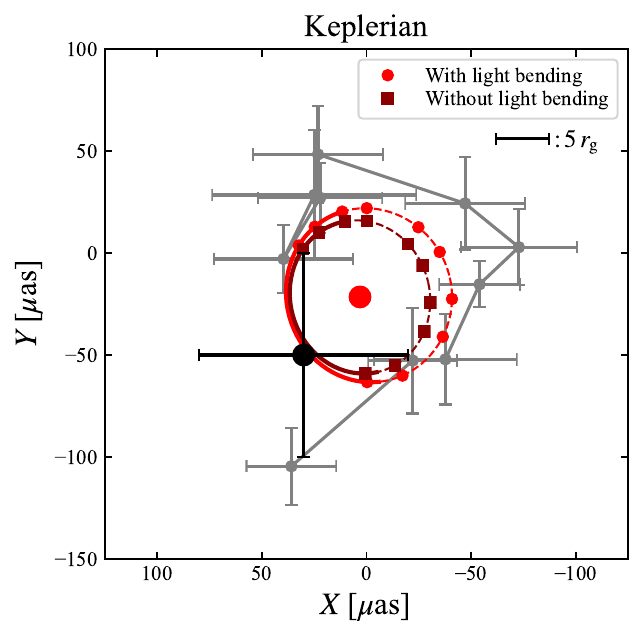}
\caption{Illustration of the impact of light bending on the appearance of an orbit on the sky. The red curve is identical to the one in Panel (a) of Fig.~\ref{fig fit}. The dark red curve depicts the same orbit projected onto the sky without accounting for light bending.
Only the half of the orbit further away from the observer, shown here dashed, is significantly affected by light bending because light from that half grazes past the BH on its way to the observer.}
\label{fig lightbending}
\end{center}
\end{figure}

To illustrate the impact of light bending on the fit, Fig.~\ref{fig lightbending} depicts the apparent hotspot orbits with and without light bending. Even for a small orbit of radius $r_0 = 7.5\,r_{\rm g}$, light bending is significant only for the half of the orbit behind the BH. Light from the half between the BH and the observer is largely undeflected.

The quality of the fit is not improved significantly by allowing the BH to have nonzero spin (see also G18).
The Keplerian orbital period at radius $r$ on the equatorial plane of a spinning BH is $P_{\rm K}=2\pi(r_{\rm g}/c)[(r/r_{\rm g})^{3/2}+a]$, where $|a|\le1$ is the dimensionless spin parameter of the BH \citep{Bardeen+1972}.
For the same orbital period, as $a$ decreases from 0 to $-0.5$, the radius of the orbit rises only slightly from $7.5\,r_{\rm g}$ to $7.6\,r_{\rm g}$, and $a$ cannot be any smaller because the orbit would then be inside the innermost stable circular orbit (ISCO).

Another way spin can affect the fit is through the slightly different light-bending effects.
However, light bending is of secondary importance if the best-fit orbit for any spin, like the zero-spin fit above, is both large and face-on (see G18).
The best-fit trajectories found in the following sections are even larger, so light bending plays an even smaller role.

\subsubsection{Geodesic}
\begin{figure}
\begin{center}
\includegraphics[width=85mm, angle=0]{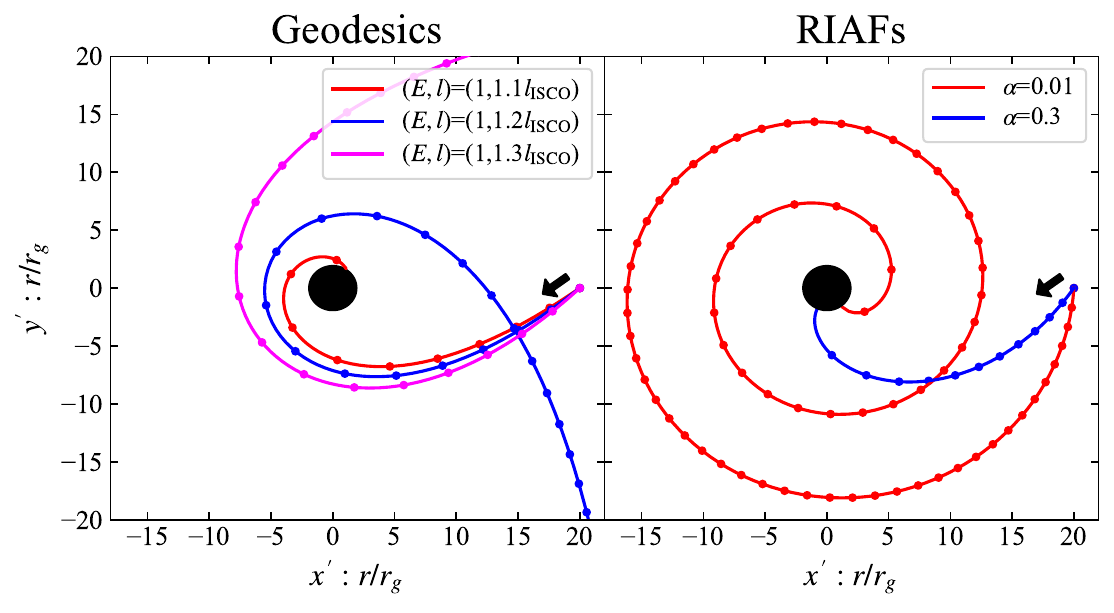}
\caption{Non-circular trajectories in our models viewed face-on ($i=0^\circ$) without light bending.
The dots on each curve are marked every $\delta t=10\,r_{\rm g}/c$; for comparison, the observed flare duration is $t\simeq90\,r_{\rm g}/c$. ({\bf Left}) Geodesics: The three marginally bound ($E=1$) geodesics have different angular momenta: $l/l_\mathrm{ISCO}=1.1$ (red), 1.2 (blue), and 1.3 (magenta). Only the first geodesic plunges. The $l/l_{\rm ISCO}=1.2$ (or $l/(r_{\rm g}c)=4.16$) geodesic is approximately the best-fit $E=1$ geodesic (see Fig.~\ref{fig el}).
({\bf Right}) RIAF streamlines \citep{Popham&Gammie1998b}: The red and blue streamlines have viscosity parameters $\alpha=0.01$ and 0.3, respectively.
}
\label{fig orbit}
\end{center}
\end{figure}

Geodesics  generalize the circular Keplerian orbits considered in \S \ref{sec kepler}. They are the most general orbits in the absence of magnetic, pressure, and other
non-gravitational forces.
A Schwarzschild geodesic is characterized by $E$ and $l$, its specific energy and specific angular momentum  \citep[e.g.,][]{Shapiro&Teukolsky1983}.
A particle with  $E<1$ ($>1$) is bound to (unbound from) the BH.
Marginally bound geodesics with $E=1$ and $l/l_{\rm ISCO}\in\{1.1,1.2,1.3\}$ are shown in the left panel of Fig.~\ref{fig orbit}, where $l_{\rm ISCO}=2\sqrt{3}\,r_{\rm g}c\simeq3.46\,r_{\rm g}c$ is the specific angular momentum of a particle orbiting at the ISCO.

The motion of the hotspot on the sky suggests that it makes a turn around the BH before shooting off to $\gtrsim20\,r_{\rm g}$;
therefore, we restrict our attention to initially inward-moving but non-plunging geodesics.
For non-plunging geodesics, $E$ must satisfy
\begin{align}
\label{eq condition circular orbit}
V_{\rm eff}(r_{\rm c,+};\,l) \leq E^2 \leq V_{\rm eff}(r_{\rm c,-};\,l),
\end{align}
where $V_{\rm eff}(r;l)=\big[1-{2}/{(r/r_{\rm g})}\big]\big[1+{(l/r_{\rm g}c)^2}/{(r/r_{\rm g})^2}\big]$ is the effective potential, and $r_{\rm c,\pm}$ are the radii of the outer (stable) and inner (unstable) circular orbits with specific angular momentum $l$.

\begin{figure}
\begin{center}
\includegraphics[width=75mm, angle=0]{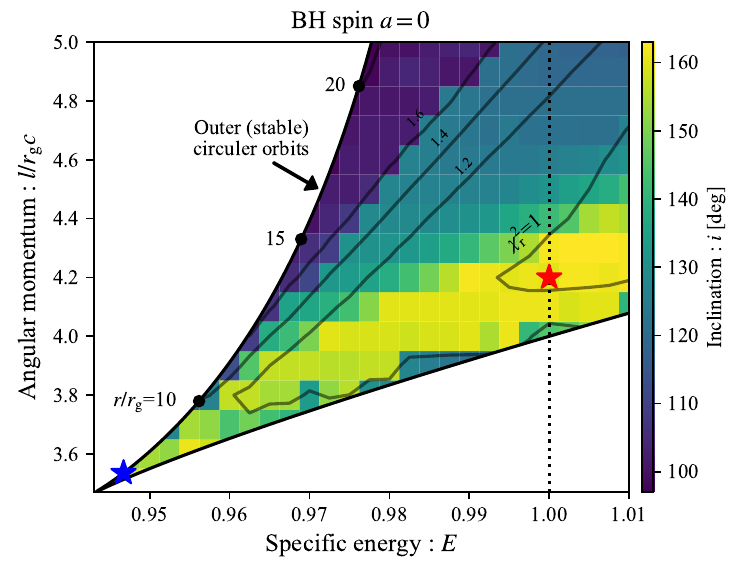}
\caption{Inclination (colors) and reduced $\chi^2$ (contours) of the best-fit geodesic in Schwarzschild spacetime as functions of its specific energy $E$ and specific angular momentum $l$.
The best-fit $E=1$ geodesic is marked with a red star; it has $l/(r_{\rm g}c)=4.2$ and $i=162^\circ$.
The best-fit circular Keplerian orbit, with $r=7.5\,r_{\rm g}$ and $(E,\,l/(r_{\rm g}c))=(0.947,\,3.53)$, is marked with a blue star.
While geodesics with $E>1$ has smaller $\chi_{\rm r}^2$, they are unrealistic because they have relativistic speeds at infinity.}
\label{fig el}
\end{center}
\end{figure}

\begin{figure}
\begin{center}
\includegraphics[width=75mm, angle=0]{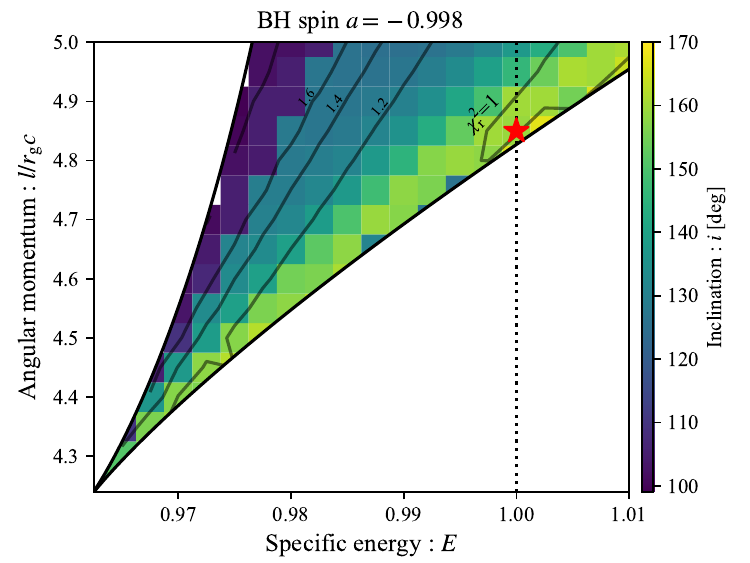}
\caption{Same as Fig.~\ref{fig el} but for a BH spin of $a=-0.998$.}
\label{fig el-1}
\end{center}   
\begin{center}
\includegraphics[width=75mm, angle=0]{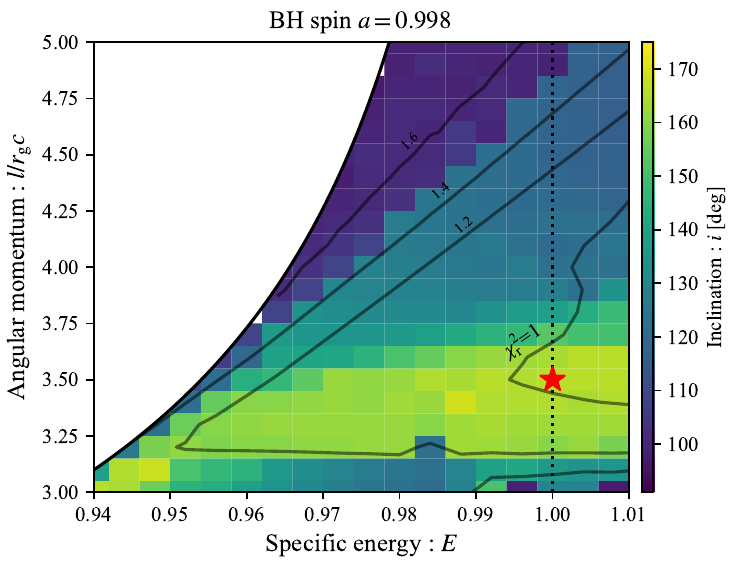}
\caption{Same as Fig.~\ref{fig el} but for a BH spin of $a=0.998$.}
\label{fig el+1}
\end{center}
\end{figure}

There are eight free parameters in this model: the six common to all models, plus $(E,l)$.
We perform a fit for each $(E,l)$ within the region demarcated by Eq.~\eqref{eq condition circular orbit}. Fig.~\ref{fig el} shows the best-fit $i$ and $\chi_\mathrm r^2$ as functions of $E$ and $l$. The value of $\chi_\mathrm r^2$ does not vary perceptibly across $E=1$; thus, we cannot tell whether marginally unbound geodesics are favored over marginally bound ones, or vice versa.  
The best-fit circular Keplerian orbit, discussed earlier in \S\ref{sec kepler} and marked with a blue star in Fig.~\ref{fig el}, has $(E,l/(r_{\rm g}c))\simeq(0.95,3.53)$ and a radius of $r=7.5\,r_{\rm g}$.

The best-fit geodesic has $(E,l/(r_{\rm g}c))=(1.01,4.3)$. It is right at the edge of the parameter space we explored. If the hotspot were due to an object falling in from large distances, it cannot conceivably follow a geodesic whose energy is very different from $E=1$; therefore, we do not consider more energetic geodesics that nominally provide a better fit.
Instead, we focus on $E=1$ geodesics; the best-fit geodesic ($l/(r_{\rm g}c)=4.2$) is marked with a red star in Fig.~\ref{fig el}, and its trajectory on the sky is illustrated in Panel (b) of Fig.~\ref{fig fit}.
The fit is still not ideal in the sense that the trajectory lies somewhat interior to the observed hotspot locations.
At lower energies, highly eccentric geodesics are better fits than circular ones. 

We can generalize the discussion above to equatorial geodesics in Kerr spacetime, which are also characterized by $E$ and $l$.
Figs.~\ref{fig el-1} and~\ref{fig el+1} depict $i$ and $\chi_{\rm r}^2$ of the best-fit geodesics around maximally spinning BHs as functions of their $E$ and $l$.
The results are qualitatively the same as in the Schwarzschild case.
The best-fit geodesics are unbound, with inclinations comparable to before.
Because the best-fit geodesics do not come close to the BH, BH spin does not significantly affect the results.
However, the best-fit geodesics are still too small compared to the observed trajectory.

These geodesics are much more eccentric than the orbits of other known objects in the Galactic Center, e.g. G2 \citep{Gillessen+2012}.
The origin of the object on such geodesics is not clear.
The cold disk \citep{Murchikova+2019}, of size $0.004\,\rm pc$, is the only structure around Sgr A* with a similar inclination as the best-fit geodesic, so the object may have come from there.
However, a geodesic approaching this close to the BH must have an angular momentum much lower than other objects in the cold disk. 
One possible candidate for the object is an asteroid tidally disrupted by the BH \citep{Zubovas+2012}, but there is no convincing mechanism by which an object on such a geodesic converts its kinetic energy into a flare of the magnitude observed.

\subsubsection{Radiatively inefficient accretion flow}\label{sec adaf}
The accretion rate onto the BH of Sgr A* is sub-Eddington \citep{Baganoff+2003,Marrone+2007}.
The BH is likely surrounded by a RIAF \citep{Yuan&Narayan2014}, whose gas is so dilute that it cannot radiate away a sizable fraction of the dissipated energy within the short amount of time it takes to fall into the BH.
The accretion flow has finite radial velocity, and sub-Keplerian azimuthal velocity because of an outward pressure gradient \citep{Narayan&Yi1994,Popham&Gammie1998b,Manmoto2000}.
A hotspot advected along the streamline of such a flow moves on an inward spiral.      
We obtain RIAF streamlines by integrating along the velocity fields calculated by \cite{Popham&Gammie1998b}. They calculated the vertically integrated structure of axisymmetric RIAFs for different viscosity parameters $\alpha$, fractions of the dissipated energy that is advected into the BH $f$, and adiabatic indices $\gamma$. The velocity fields we use pertain to $\alpha\in\{0.01,0.3\}$, $f=1$, and $\gamma=1.4444$.
The right panel of Fig.~\ref{fig orbit} depicts these streamlines. Since $\alpha$ governs the rate of outward angular momentum transport, a larger $\alpha$ makes the radial velocity larger and the azimuthal velocity smaller, which results in the streamline being more loosely wound. 
The RIAF model has the same six free parameters common to all models, while $\alpha$ is fixed to the two values above.

Panel (c) of Fig.~\ref{fig fit} depicts how the best-fit streamlines for the two values of $\alpha$ look on the sky.
The $\alpha=0.3$ streamline has a deformed image due to light bending (most of the orbit is on the far side of the BH). Its $\chi^2_{\rm r}$ is worse than the circular Keplerian model. Since the $\alpha=0.01$ streamline mimics a circular Keplerian orbit over short time intervals (see the right panel of Fig.~\ref{fig orbit}), its best-fit parameters are similar to those of the Keplerian fit. Its $\chi^2_{\rm r}$ is  comparable but slightly larger (see Fig.~\ref{fig fit}). 
This agreement provides further physical motivation for considering the circular Keplerian fit in the first place.
However, in both cases the best-fit hotspot locations all lie nearer to the BH than the observed ones, casting doubt on the validity of this model.

\subsubsection{Summary of material models}
None of models considered so far, in which a hotspot travels along a physically motivated orbit, gives a satisfactory fit.
While $\chi^2_{\rm r}$ values are reasonable in all cases, the best-fit hotspot locations are all nearer to the BH than the observed ones. 
This problem arises from the tension mentioned earlier between the short duration of the orbit and the spread of the hotspot locations on the sky. 
This motivates us to consider in the following section other models in which the hotspot is a pattern rather than a physical entity.

\subsection{Pattern models}
The difficulty in simultaneously fitting the size and duration of the observed hotspot trajectory with the orbits considered in \S\ref{sec material model} can be resolved if the hotspot moves at super-Keplerian speeds. Such speeds can arise if additional, non-gravitational forces act on a physical entity. Although the velocities required for a good fit are subluminal, they are so large that the hotspot is more likely a pattern.

Spiral patterns may be common in accretion flows \citep[see e.g.][]{Tagger&Melia2006,Falanga+2007}.
An extended spiral is inconsistent with the observations because the observed diameter of the hotspot is $\lesssim\,5\,r_{\rm g}$ (G20).
Furthermore, the flux centroid of a multi-armed spiral should not drift significantly away from the BH.
It is also inconsistent with the fact that the hotspot was observed for less than a full orbit, considering that it may take a few orbits for a spiral to develop.
Therefore, we consider here a phenomenological single point-like pattern.

\subsubsection{Super-Keplerian pattern}
\label{sec super kepler}
We consider circular motion around the BH given by Eqs. \eqref{eq kepler x} and \eqref{eq kepler y}, with $\omega_{\rm K}$ is replaced by a free parameter $\omega$.
Together with the six parameters common to all models, the total number of free parameters in this model is seven.

Panel (d) of Fig.~\ref{fig fit} depicts the best-fit trajectory.
Unlike the circular Keplerian fits, the trajectory passes through the observed hotspot locations.
The best-fit radius and angular velocity are $r_0=12.5\,r_{\rm g}$ and $\omega=2.7\,\omega_{\rm K}$ respectively, hence the rotational velocity is $v_\varphi=(\omega/\omega_\mathrm{K})v_\mathrm{K}\simeq0.76\,c$.

Let us consider several ways in which super-Keplerian motion may be physically realized.
One possibility is that forces other than gravity acts on the hotspot, such as the inward pressure gradient near the ISCO of an accretion disk \citep{Abramowicz+1978,Kozlowski+1978,Popham&Gammie1998b}.
However, the required pressure gradient is $\sim6$ times stronger than gravity, which is extremely unlikely.
Alternatively, \cite{Tursunov+2019} argued that a charged hotspot in the poloidal magnetic field around the BH (G18) could move at super-Keplerian speeds due to the Lorentz force.

The observed super-Keplerian motion is more likely a pattern motion, such as a wave. The cause of the pattern motion is unlikely hydrodynamic because the sound speed is at most of order Keplerian:
\begin{align}
c_{\rm s}\simeq\sqrt{\frac{k_{\rm B}T_{\rm i}}{m_{\rm p}}}\lesssim\sqrt{\frac{GM}{r}}=v_{\rm K},
\end{align}
 where $k_{\rm B}$, $T_{\rm i}$, and $m_{\rm p}$ are the Boltzmann constant, ion temperature of the accretion flow, and proton mass, respectively.
The inequality is because the ion temperature cannot exceed the virial temperature, $k_{\rm B}T_{\rm i}\lesssim GMm_{\rm p}/r$. 
However, the pattern motion could be magnetohydrodynamic because the Alfv\'en speed during flares is larger than the Keplerian orbital speed:
\begin{align}
v_\mathrm A&=\biggl(\frac{B^2}{4\pi\rho}\biggr)^{1/2}\simeq0.9\,c\,
  \biggl(\frac B{100\,\mathrm{G}}\biggr)
  \biggl(\frac\rho{10^{-18}\,\mathrm{g\,cm^{-3}}}\biggr)^{-1/2}\nonumber\\
  &> v_{\rm K}\simeq0.3\,c\,(r_0/12.5\,r_{\rm g})^{-1/2}.
\end{align}
The fiducial density is appropriate for the quiescent state \citep{Loeb&Waxman2007}, but the fiducial magnetic field is for the flaring state \citep[e.g.,][]{DoddsEden+2009}. One piece of evidence supporting the existence of strong magnetic fields in the vicinity of the BH during flares is the synchrotron nature of the NIR emission.

\subsubsection{Precessing pattern}\label{sec precession model}
So far we have considered a hotspot moving in a plane containing the BH. As one generalization, we turn to the case of hotspot motion on a different plane.
It could be a pattern motion excited on a surface by an external source (such as a precessing outflow interacting with a surrounding disk).
For simplicity, we assume the hotspot follows an arc of a circle, and the perpendicular line passing through the center of the circle contains the BH; the most general hotspot motion can be much more complicated.
Our goal is to estimate the radius and inclination of the circle.
We repurpose the super-Keplerian machinery (\S\ref{sec super kepler}) to fit the hotspot trajectory, except that we ignore light bending here because we assume the entire hotspot trajectory lies between the observer and the BH for simplicity (see Fig.~\ref{fig lightbending}).
The projection of the hotspot trajectory on the sky is therefore an ellipse.
Given some best-fit circular trajectory, one may expect that the same trajectory rotated $180^\circ$ about the observer's line of sight would be an equally good fit, but the effect of light travel time breaks this symmetry.

Panel (e) of Fig.~\ref{fig fit} depicts the  best-fit hotspot trajectory and the axis through the center of the circle. The radius of the trajectory is $r_0=14.1\,r_{\rm g}$, and the hotspot moves around the circular trajectory at $v_\varphi\simeq0.88\,c$.
The BH can be anywhere along the axis, as long as it lies inside the error box of the location of Sgr A* constrained by the orbital motion of S2.
The furthest the BH can be from the center of the circle is at the north-east end of the axis; at this point, the BH is $\simeq16.2\,r_\mathrm g$ from the center of the circle, the hotspot is $\sim20\,r_\mathrm g$ from the BH, and the hotspot subtends an angle $\sim40^\circ$ from the axis at the BH. 
Comparing our results here with our super-Keplerian fit (\S\ref{sec super kepler}), we find that the best-fit radius increases only slightly when the BH is allowed to be off the plane of the hotspot trajectory.

\section{Discussion and Summary}\label{sec summary}
G18 observed the 2018 July 22 flare of Sgr A* with the NIR interferometry instrument GRAVITY. They measured the motion of a hotspot with an unprecedented level of astrometric accuracy ($\sim10\,\muup\mathrm{as}$). This allows us, for the first time, to study dynamics close to the BH. 
The hotspot traced out a partial loop on the sky within a time shorter than what circular Keplerian orbits would predict.
Circular Keplerian fits (G18, G20, see also \S\ref{sec kepler}) result in best-fit hotspot locations that are nearer to the BH than the observed ones.\footnote{We approximated here the hotspot as a point source. While the consideration of an extended source could in principle alleviate the problem, G20 estimated that the diameter of the source should be $\lesssim5\,r_\mathrm g$.} This motivates us to explore a broader range of hotspot models.

As a first step, we consider several kinematic models in which the hotspot is confined to a plane.
Our models are categorized into two types: material models, whose hotspot tracks some fluid element; and pattern models, whose hotspot follows some pattern motion.
The material models include the circular Keplerian model, the geodesic model, and the RIAF  model; the last one is based on models of the accretion flow onto Sgr A*.
The pattern models include the super-Keplerian pattern model and the precessing pattern model.

All material models suffer from the same problem: The size of the best-fit trajectory is only $r\simeq 7.5\,r_\mathrm g$, largely because it is dictated by the short time the hotspot took to traverse the trajectory. The material models fail to reproduce the observed trajectory size on the sky, even though the fits appear to be satisfactory as judged by the reduced $\chi^2$ metric. 
Although our model fits assume a zero-spin BH, the situation is not improved for nonzero spin. Two other flares not studied here, the May 27 and July 28 flares, are also similar to the July 22 flare in that the hotspot moves across a large angle on the sky in a short time.

The pattern models do much better because the trajectory size is decoupled from the traversal time.
The best-fit trajectory size in  these models is $\gtrsim 12\,r_\mathrm g$.
This is larger than the ISCO radius at any spin, in contradiction with the expectation that the hotspot should be produced by the innermost parts of some accretion disk; therefore, we cannot constrain the BH spin based on this expectation.

Detailed comments on individual models follow:
\begin{enumerate}
\item We confirm the results of G18 and G20, namely, the best-fit circular Keplerian orbit has a small radius $r\simeq 7.5\,r_\mathrm g$ that is very close to the ISCO of a Schwarzschild BH.
Our fit suffers from the same problem that the best-fit hotspot locations are all nearer to the center of the orbit than the observed ones (see also G20), suggesting that circular Keplerian orbits are poor fits.
A spinning BH does not change the period of the orbit so much that it improves the fit significantly. 
If the best-fit orbit is indeed the orbit of the hotspot, its size constrains the BH spin to $>-0.5$.
Turning this argument around, because the hotspot is expected to be near the ISCO, the best-fit orbit, which is close to the Schwarzschild ISCO, suggests that the BH is slowly rotating.
However, as mentioned earlier, we find this fit problematic.

\item A hotspot advected along a RIAF streamline gives us as good a fit as a circular Keplerian orbit when outward angular momentum transport is inefficient ($\alpha=0.01$). A larger $\alpha$ leads to faster inspiral and worse fit. 

\item A marginally bound geodesic in Schwarzschild spacetime provides a slightly better fit than a circular Keplerian orbit, but even in this case, the best-fit geodesic still lies inside the observed hotspot locations. The results are qualitatively the same if we consider instead equatorial geodesics in Kerr spacetime.
It is unclear where the object responsible for the hotspot originated from.
Among known Galactic Center objects, the cold disk \citep{Murchikova+2019} of size $0.004\,\rm pc$ is the only structure with a similar inclination as the best-fit geodesic.
It is also unclear how the kinetic energy of the object falling toward the BH is converted to a flare of the kind observed here.

\item A satisfactory fit can be obtained with a hotspot moving in a circle around the BH
at $r\simeq 12.5 r_{\rm g}$ and at a velocity 2.7 times super-Keplerian, but no known force could compel an object to go around the circle at such speeds.
The super-Keplerian motion could instead be the pattern motion of a magnetohydrodynamic perturbation in a strongly magnetized gas surrounding the BH.

\item An equally good model is a hotspot traveling at super-Keplerian speeds along a circle on a plane offset from the BH.
This could be the pattern motion created by a precessing outflow interacting with a surrounding disk.
The furthest the plane can be from the BH is $\simeq16.2\,r_{\rm g}$; in this case, the distance of the hotspot from the BH is $\sim20\,r_{\rm g}$.
\end{enumerate}

The last two models, which are pattern models, have $\chi_{\rm r}^2=0.6$, suggesting an ``overfit'' of the data.
This is a hint that the current data do not allow us to constrain models more complicated than our simple pattern models.
More data or better data would be needed to go beyond them.

So far our discussion is focused on the astrometry of the best-observed flare, the 2018 July 22 one. 
It is interesting to note that all other flares have the same sense of rotation around the BH. 
Future flares with similar or better astrometric accuracy would confirm whether this is generally true.
Models could then be ruled out based on whether the hotspot has a preferred sense of direction (e.g., RIAF or pattern motion) or not (e.g., geodesic).

The light curve of the July 22 flare has a single peak (G18). The
rising part of the light curve is attributed to the onset of some
unexplained mechanism giving rise to the hotspot, while the falling part
is related to the lifetime of the hotspot, limited by differential
rotation and synchrotron cooling. Analysis of the light curve is beyond
the scope of this paper.
The observation of a single spot suggests that, regardless of the mechanism that produced the pattern, the decay is fast enough that a spiral structure would not have time to form.

Our models are quite restrictive in their simplifying assumptions, the strongest of which could be the requirement that the hotspot be a point source confined to a plane. Clearly, the shape and trajectory of the hotspot can both be much more complicated; for example, the trajectory could be an outgoing spiral along a twisted magnetic field line of an outflow.
Furthermore, we have explored only the kinematics here.
The shape and trajectory of the hotspot must ultimately be derived from magnetohydrodynamics considerations.
This is true regardless of the provenance of the hotspot: a free-falling object tearing through some ambient medium, perturbations in a RIAF, a precessing outflow interacting with a surrounding disk, clumps ejected from the BH along twisted magnetic field lines, or other more exotic scenarios that we have not explored here. General relativistic magnetohydrodynamics simulations are invaluable for modeling these hotspots.
These simulations would also open up explorations of other aspects of flares, such as how the flux and polarization angle vary as functions of time.

\section*{acknowledgments}
We thank Julian Krolik and Re'em Sari for fruitful discussions, and an anonymous referee for helpful comments.
This work is supported in part by JSPS Postdoctral Fellowship, Kakenhi No. 19J00214 (T.M.) and by ERC advanced grant ``TReX'' (C.H.C. and T.P.).

\section*{data avialbility}
The data underlying this article will be shared on reasonable request to the corresponding author.

\bibliographystyle{mnras}
\bibliography{reference_matsumoto}

\bsp
\label{lastpage}
\end{document}